# Effective quantum number for axially symmetric problems

N.N.Trunov[1]

*D.I.Mendeleyev Institute for Metrology*

*Russia, St.Petersburg. 190005 Moskovsky pr. 19*

*trunov@vniim.ru*

(Dated: August 15, 2014)

**Abstract.** We generalize the universal effective quantum number introduced earlier for centrally symmetric problems. The proposed number determines the semiclassical quantization condition for axially symmetric potentials.

## 1. Introduction

We have earlier introduced and calculated a new universal effective quantum number which determines the level ordering and other properties of the centrally symmetric systems, though it is not formally an exact quantum number[1]. The quantization condition may be written as

$$\Phi(E) = \frac{\sqrt{2m}}{\pi\hbar} \int \sqrt{E-V}\, dr = T \quad . \tag{1}$$

with a new effective quantum number $T$ including radial $n$ and orbital $l$ quantum numbers

$$T(n,l) = (n + \tfrac{1}{2}) + \varphi(l + \tfrac{1}{2}) \tag{2}$$

and the functional $\varphi$,

---
[1] Electronic address: trunov@vniim.ru

$$\varphi^2 = \frac{1}{2\pi^2} \frac{\left[\int dr \sqrt{2(E-V)}\right]^3}{\int r^2 dr \left[2(E-V)\right]^{3/2}} . \qquad (3)$$

Here $V$ is a given potential without the centrifugal one, $E$ - the energy. The value of $\varphi$ does not depend on $E$ for power-law potentials and is a smooth function of $E$ for other ones.

We also have generalized the above approach for a special case of an axial symmetry, namely for an elliptic deformation of a given potential [2]:

$$V(r) \to V(s), \qquad (4)$$
$$s^2 = (x^2 + y^2)/a^2 + z^2/c^2, \qquad (5)$$
$$a^2 c = 1. \qquad (6)$$

In the present paper we treat arbitrary axially symmetric potentials with a moderate difference from a centrally symmetric one.

## 2. Intermediate calculations

Introducing a small parameter α such that
$$a = 1 - \alpha/3, \quad c = 1 + 2\alpha/3, \qquad (7)$$
and $\varsigma = \cos\theta = z/r$, we get our new potential
$$W(r, \varsigma) = V(s),$$
$$s = r(1 - 2\alpha P_2(\varsigma)). \qquad (8)$$
Here and below we use linear in α approximation; $P_n$ are the Legendre polynomials.

The common phase integral depends now on $\varsigma$:
$$I(\varsigma) = \int dr \sqrt{\varepsilon - W(r,\varsigma)} = I_1(1 + 2\alpha P_2(\varsigma)), \qquad (9)$$

$$I_1 = \int du \sqrt{\varepsilon - V(u)}. \tag{10}$$

Taking into account properties of $P_n$, we have

$$I_1 = \frac{1}{2}\int I(\varsigma)d\varsigma, \tag{11}$$

$$\alpha = \frac{15}{4I_1}\int P_2(\varsigma)I(\varsigma)d\varsigma. \tag{12}$$

Note that the same result we obtain as the mean-square approximation with two parameters, $I_1$ and $\alpha$:

$$\Delta = \int [I(\varsigma) - I_1(1 + 2\alpha P_2(\varsigma))]^2 \, d\varsigma = \min \tag{13}$$

Thus we can use (11) and (12) as the best approximation for any axially symmetric potential $W$ with a moderate deviation from a centrally symmetric one (certainly $\Delta = 0$ for an elliptic deformation (4-6)).

At last, the integral in the denominator (3) appears as the total number of eigenstates with energies not exceeding $E$. For our potentials W it looks like

$$I_3 = \frac{1}{2}\int d\varsigma \int r^2 dr (E - W)^{3/2} \tag{14}$$

## 3. Effective quantum number

For a special case of an elliptic deformation (4), the corresponding expression is [2]:

$$T(n,l,m) = T(n,l)(1 + 2\alpha f(m,l)) \tag{15}$$

with $T(n,l)$ from (1), $m$ – the magnetic quantum number and

$$f(m,l) = \frac{1}{(2l-1)(2l+3)}\left[m^2 - \frac{l(l+1)}{3}\right]. \tag{16}$$

In our general case $W(r,\varsigma)$ we have to use $\alpha$ (12). The quantization condition takes the following form:

$$\frac{\sqrt{2m}}{\pi\hbar}I_1(E) = T(n,l,m) \tag{17}$$

with $I_1$ (11) and $T$ (15), where

$$\varphi^2 = \frac{1}{2\pi^2} \frac{I_1^3}{I_3^{3/2}} \qquad (18)$$

It coincides certainly with (1) - (3) for centrally symmetric problems.

Thus we have generalized the effective number approach for axially symmetric potentials with a moderate difference from centrally symmetric one.

1. L.A.Lobashev and N.N.Trunov. A universal effective quantum number for centrally symmetric problems. J.Phys.A: Math. Theor. <u>42</u> (2009) 345202 (15pp).
2. N.N.Trunov. Effective quantum number for describing nanosystems. arXiv: 1307.4295 [quant-ph].